\documentclass{svproc}
\usepackage{url}

\usepackage{amssymb,amsmath}

\usepackage { graphicx } 

\graphicspath{{./Figures/}}

\usepackage{color} 

\usepackage{hyperref}
\hypersetup{
    colorlinks=true,
    linkcolor=blue,
    filecolor=blue,      
    urlcolor=blue,
    citecolor=blue}
                       
\def\iid{\overset{\textnormal{iid}}{\sim}} 

\def\N{\mathbb{N}}\def\R{\mathbb{R}}

\def\Lcal{\mathcal{L}}\def\Ocal{\mathcal{O}}

\begin{document}
\mainmatter              
\title{Bayesian Inference for Initial Heat States with Gaussian Series Priors}
\titlerunning{Bayesian Inference for Initial Heat States}  
%
\author{Matteo Giordano}
\authorrunning{Matteo Giordano} 
%
\tocauthor{Mateo Giordano}
\institute{ESOMAS Department, University of Turin, Corso Unione Sovietica 28/bis, 10134 Turin, Italy,\\
\email{matteo.giordano@unito.it}
}

\maketitle              

\begin{abstract}
We consider the statistical linear inverse problem of recovering the unknown initial heat state from noisy interior measurements over an inhomogeneous domain of the solution to the heat equation at a fixed time instant. We employ nonparametric Bayesian procedures with Gaussian series priors defined on the Dirichlet-Laplacian eigenbasis, yielding convenient conjugate posterior distributions with explicit expressions for posterior inference. We review recent theoretical results that provide asymptotic performance guarantees (in the large sample size limit) for the resulting posterior-based point estimation and uncertainty quantification. We further provide an implementation of the approach, and illustrate it via a numerical simulation study.
\end{abstract}

%
%
%
%
%

\section{Introduction}

Partial differential equations (PDEs) are ubiquitous mathematical models that describe the behaviour of a vast number of physical systems in engineering and the applied sciences. The formulation of PDEs typically involves one or more functional parameters, which are often unknown and not directly accessible to measurements. In order to calibrate a PDE model and use it in practice, its parameters need to be estimated beforehand, giving rise to an \textit{inverse problem} of \textit{parameter identification}. This class of problems has widespread applications in -- among the others -- imaging, geophysics, acoustics and finance, and has been extensively studied within the statistics and applied mathematics communities; see e.g.~the monographs \cite{EHN96,KS04,kaltenbacher2005parameter,AMOS19,N23}, and the many references therein.

	In this article, we shall focus on the specific problem of \textit{inverse heat conduction}, consisting in the recovery of the initial condition of the heat equation from observations of the heat distribution at a later time instant. More in details, let $\mathcal{O}\subset\mathbb{R}^d, \ d\in\N$, be a non-empty, bounded, smooth and convex set. We are interested in estimating the unknown (sufficiently regular) initial condition $f:\Ocal\to\R$ in the parabolic PDE
\begin{equation}
\label{Eq:PDE}
\begin{cases}
	\partial_t u - \nabla\cdot(c \nabla u) = 0, & (t,x)\in(0,\infty)\times\Ocal, \\
	u=0, & (t,x)\in(0,\infty)\times\partial\Ocal,\\
	u(0,x)=f(x), & x\in\Ocal,
\end{cases}
\end{equation}
where $c :\Ocal\to(0,\infty)$ is a known smooth \textit{conductivity function}, from noisy measurements of the PDE solution $Gf\equiv u(T,\cdot) :\Ocal\to\R$ at time $T>0$ over interior design points,
\begin{equation}
\label{Eq:Obs}
	Y_i = Gf(x_i) +   \sigma W_i, 
	\quad x_i\in\Ocal,
	\quad \sigma>0,
	\quad W_i\iid N(0,1),
	\quad i=1,\dots,n.
\end{equation}

\begin{figure}[t]
\label{Fig:StatProb}
\centering
\includegraphics[width=3.75cm,height=3.25cm]{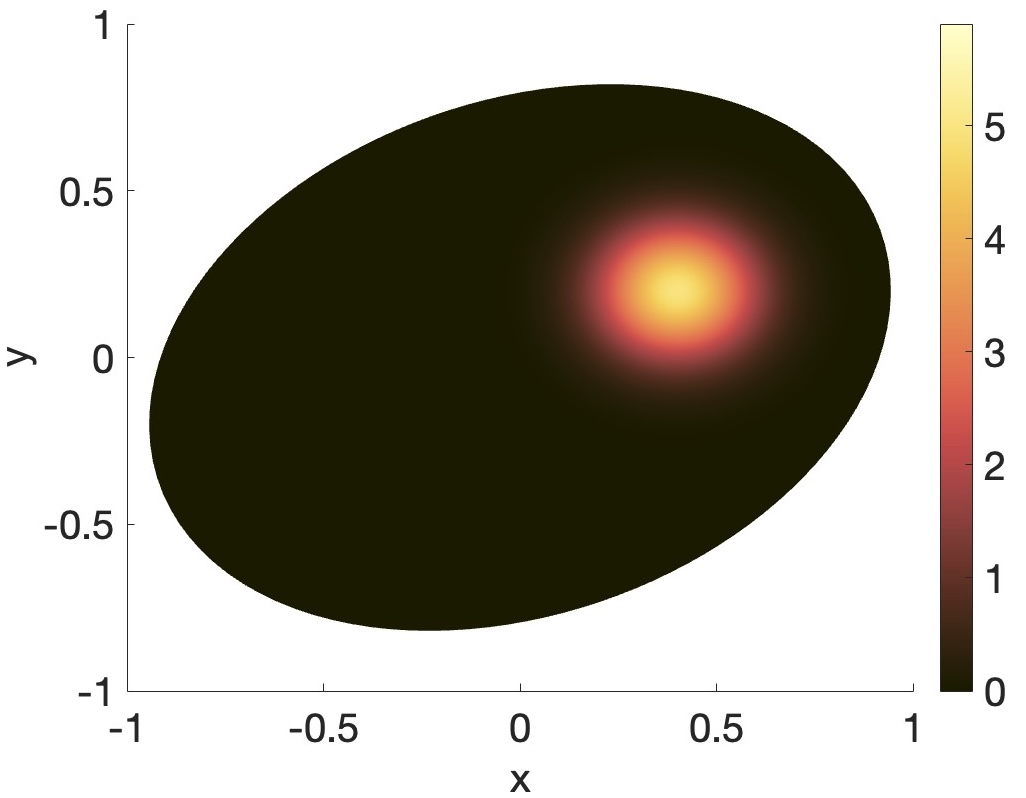}
\includegraphics[width=3.75cm,height=3.25cm]{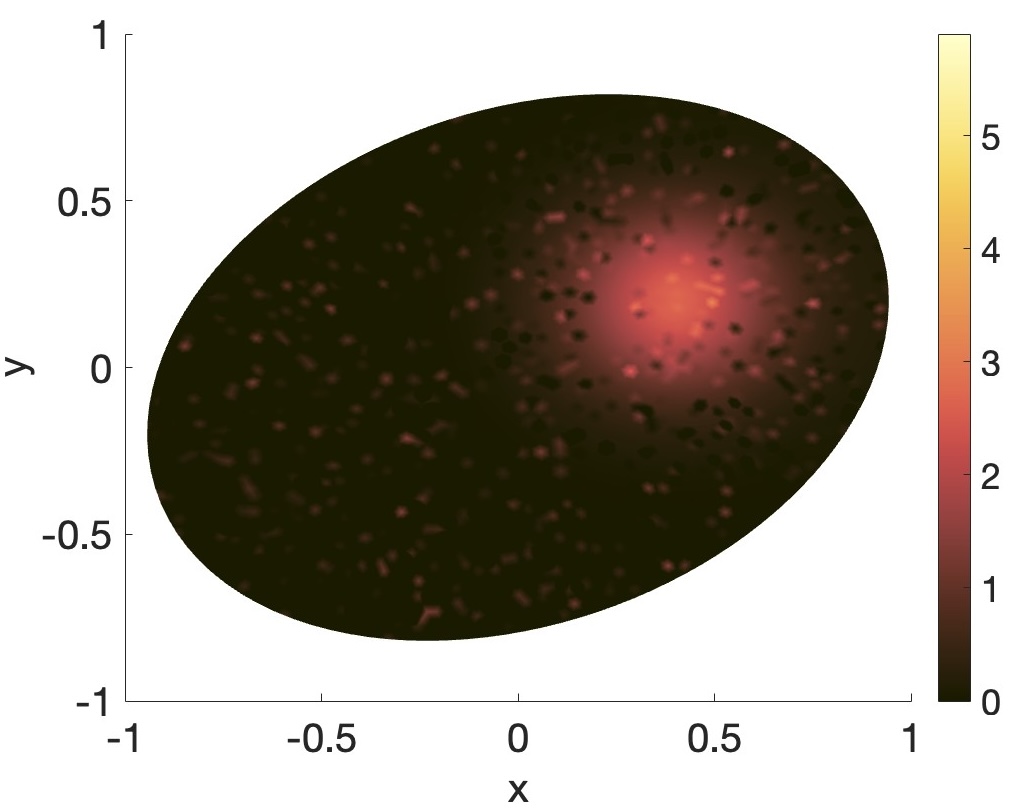}
\caption{Left: the ground truth initial condition $f_0$, defined in \eqref{Eq:f0}. Right: $n=1000$ noisy observations from the corresponding solution to the heat equation at time $T=.01$, with noise level $\sigma=.05$.}
\end{figure}

An illustration of this statistical inverse problem with synthetic data is provided in Figure \ref{Fig:StatProb}. Practical applications of the recovery of initial heat states are overviewed in \cite{EHN96}. Statistical approaches to the inverse problem of heat conduction have been studied both in the frequentist (e.g.~\cite{mair1996statistical,C08,bissantz2008statistical}) and Bayesian literature (e.g.~\cite{S10,KvdVvZ13,ASZ14}). This problem is known to be \textit{(severely) ill-posed}, in that while the \textit{parameter-to-solution map} $f\mapsto G(f)$ is well-defined, linear and continuous between suitable function spaces (e.g.~\cite{E10}), it generally does not posses a continuous inverse. This implies that regularisation is necessary to achieve a solution and to address the noise corruption, and also results in slow logarithmic minimax-optimal rates of estimation \cite{bissantz2008statistical,KvdVvZ13}.

	Recently, Giordano and Kekkonen \cite{GK20} investigated the theoretical performance of nonparametric Bayesian procedures with Gaussian priors for general linear inverse problems (under an asymptotically equivalent measurement scheme), identifying abstract conditions for the prior distribution and the forward map to obtain \textit{Bernstein-von Mises theorems}. These characterise the asymptotic shape of the posterior distribution in the large sample size limit and under the frequentist assumption that the data have been generated by some fixed ground truth. For the recovery of initial heat states, their results imply that a wide class of Gaussian process priors yield valid and optimal point estimation and uncertainty quantification, via the posterior mean estimate and credible intervals centred around it.

	Here, we shall focus on the case of Gaussian series priors, employing the Dirichlet-Laplacian eigenbasis, which offers a convenient approach to implementation for general domains and dimensions. We shall argue that these priors fit within the theoretical conditions set forth in \cite{GK20}, and review the resulting asymptotic results. In view of the linearity of the \textit{forward map} $f\mapsto Gf$ and the normal assumptions on the noise,  endowing $f$ in \eqref{Eq:Obs} with a Gaussian prior distribution results in a Gaussian conjugate posterior, cf.~\eqref{Eq:Conjugacy} below. For the considered Gaussian series priors, we shall then derive the explicit formulae for the posterior mean and covariance matrix, under a natural discretisation scheme. These findings shall be complemented in Section \ref{Sec:Numerics} with a numerical simulation study, where excellent agreement with the theory is obtained. The reproducible MATLAB code used for the study is available at: \href{https://github.com/MattGiord/Bayesian-Heat-Equation}{\texttt{https://github.com/MattGiord}}.
	
%
%
%
%
%

\section{Gaussian Series Priors for Initial Heat States}

%
%
%

\subsection{Prior Specification and Gaussian Conjugate Formulae} 

Let $\{e_j, \ j\in\N\}$ be the orthonormal basis of $L^2(\Ocal)$ (the space of square-integrable functions on $\Ocal$) formed by the eigenfunctions of the (negative) Dirichlet-Laplacian:
\begin{equation}
\label{Eq:Eigen}
\begin{cases}
	-\Delta e_j =\lambda_j e_j,  & x\in \Ocal, \\
	e_j=0, & x\in \partial\Ocal,
\end{cases}
         \qquad j\in\N,
\end{equation}
with associated eigenvalues $0<\lambda_1<\lambda_2\le\lambda_3\le\dots, $ satisfying $\lambda_j = O(j^{2/d})$ as $j\to\infty$ according to Weyl's asymptotics; see e.g.~\cite[Section 7.4]{HT07} for details. For fixed $\alpha > d/2$, we endow the unknown initial heat state $f$ in \eqref{Eq:Obs} with the following Gaussian series prior
\begin{equation}
\label{Eq:Prior}
	\Pi:=\Lcal(F),
	\qquad 
	F(x):= \sum_{j=1}^\infty \lambda_j^{-\alpha/2} F_j e_j(x),
	\qquad x\in\Ocal,
	\qquad F_j\iid N(0,1).
\end{equation}

	By construction, $\Pi$ defines a centred Gaussian Borel probability measure on the ambient space $L^2(\Ocal)$. The parameter $\alpha$ in \eqref{Eq:Prior} regulates the (Sobolev-type) regularity of the realisations of $F\sim \Pi$; in particular,  $F\in H^\beta(\Ocal)$ almost surely for all $\beta<\alpha-d/2$ and its reproducing kernel Hilbert space (RKHS) is included in $H^\beta(\Ocal)$; see \cite{nickl2024consistent} for details, and \cite[Chapter 11]{GvdV17} for the necessary background on Gaussian processes and probability measures.

\begin{figure}[t]
\centering
\includegraphics[width=3.75cm,height=3.25cm]{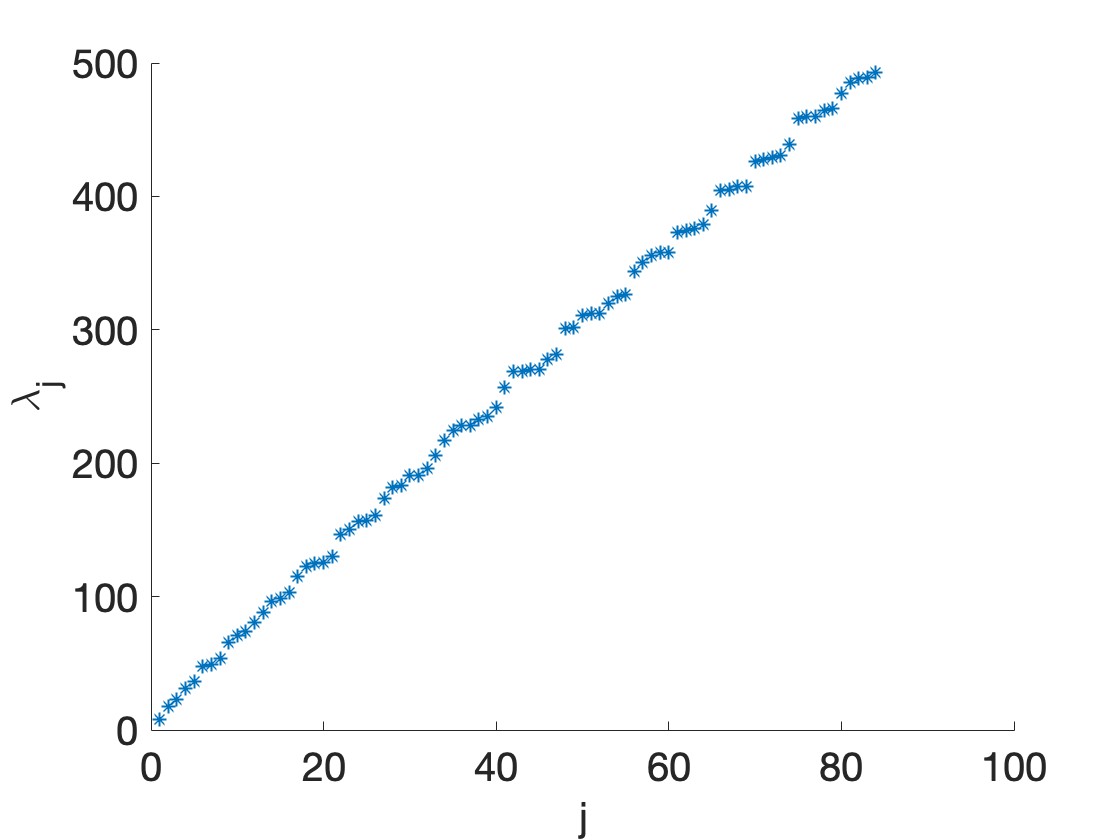}
\includegraphics[width=3.75cm,height=3.25cm]{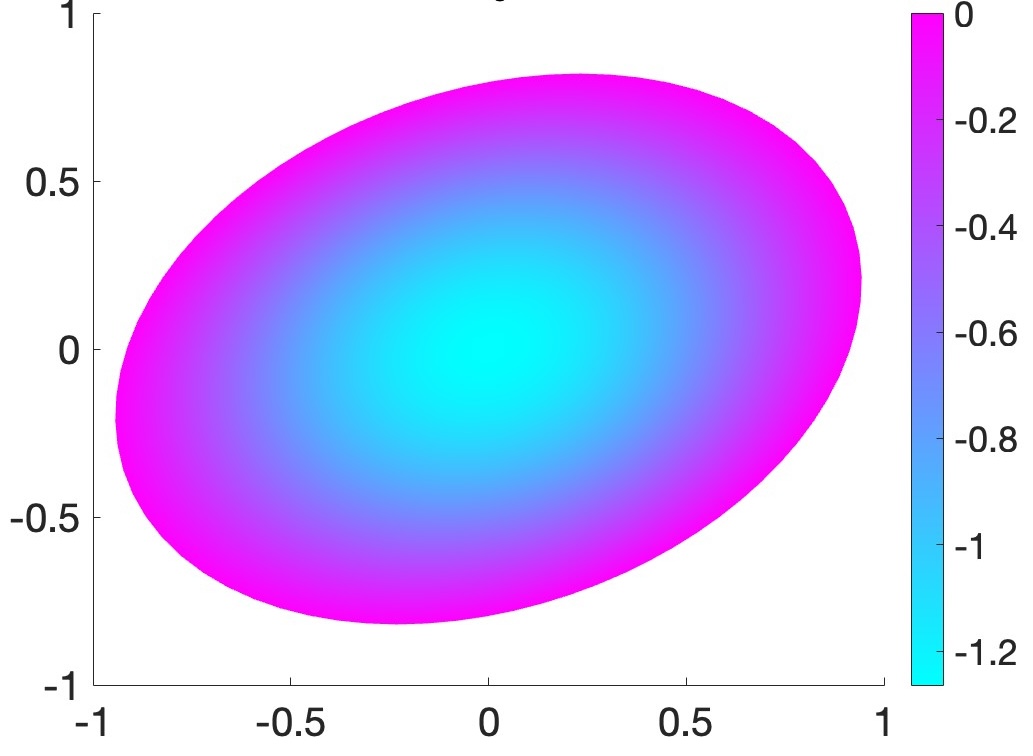}
\includegraphics[width=3.75cm,height=3.25cm]{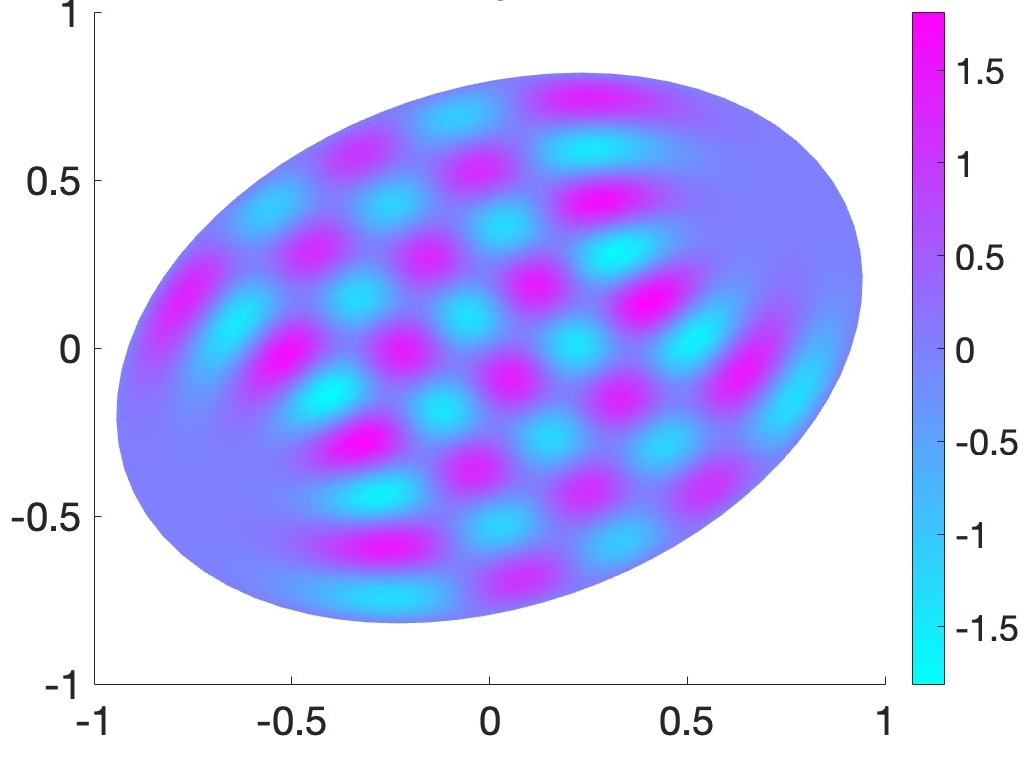}
\caption{Left: the numerical approximations to the Dirichlet-Laplacian eigenvalues for the given domain. Centre and right: numerical approximations of the first and fiftieth eigenfunction, $e_1$ and $e_{50}$, respectively.}
\label{Fig:Eigen}
\end{figure}

	In practice, outside certain special cases for the domain $\Ocal$ (such as squared or circular ones), the Dirichlet-Laplacian eigenpairs are not explicitly available. However, for general domains, they can be numerically approximated by solving the elliptic eigenvalue problem \eqref{Eq:Eigen} via off-the-shelf finite element methods available in many statistical and mathematical computing environments, see Figure \ref{Fig:Eigen}.
	
\begin{figure}[t]
\centering
\includegraphics[width=3.75cm,height=3.25cm]{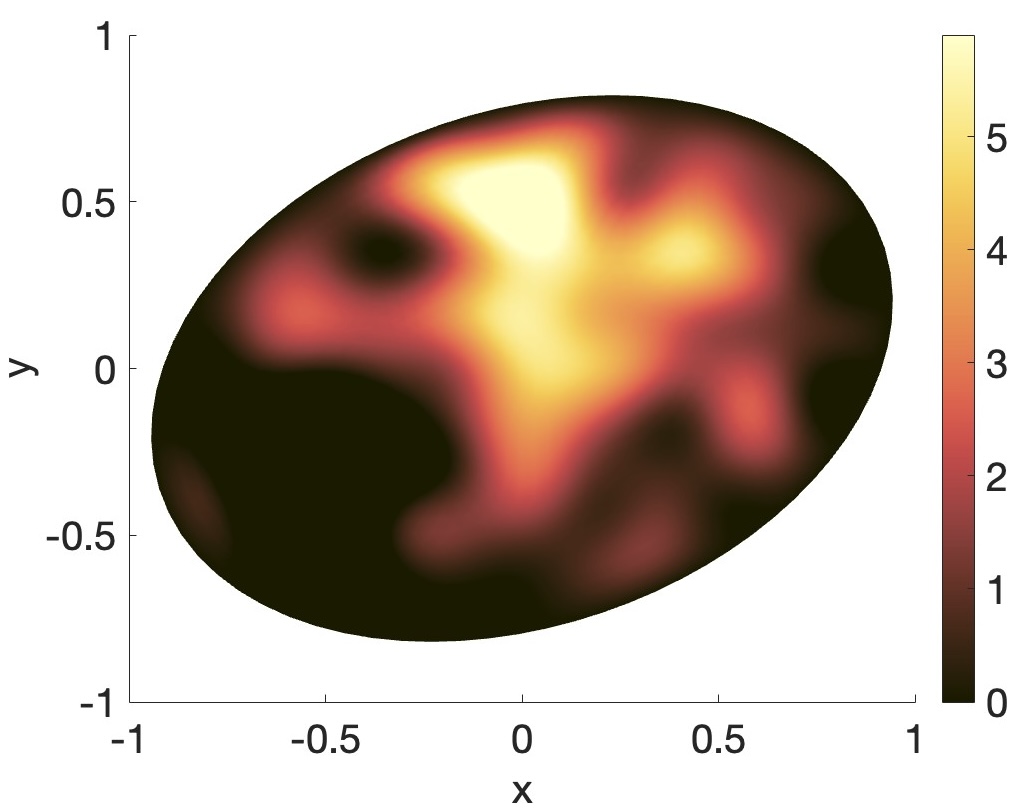}
\includegraphics[width=3.75cm,height=3.25cm]{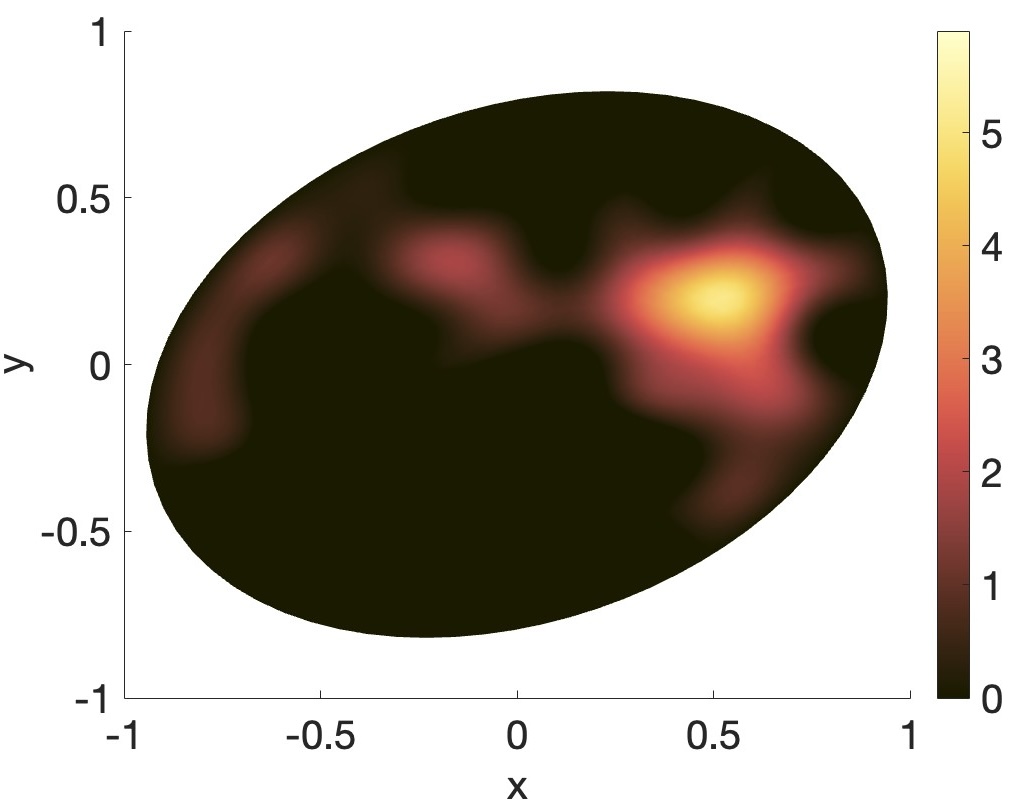}
\includegraphics[width=3.75cm,height=3.25cm]{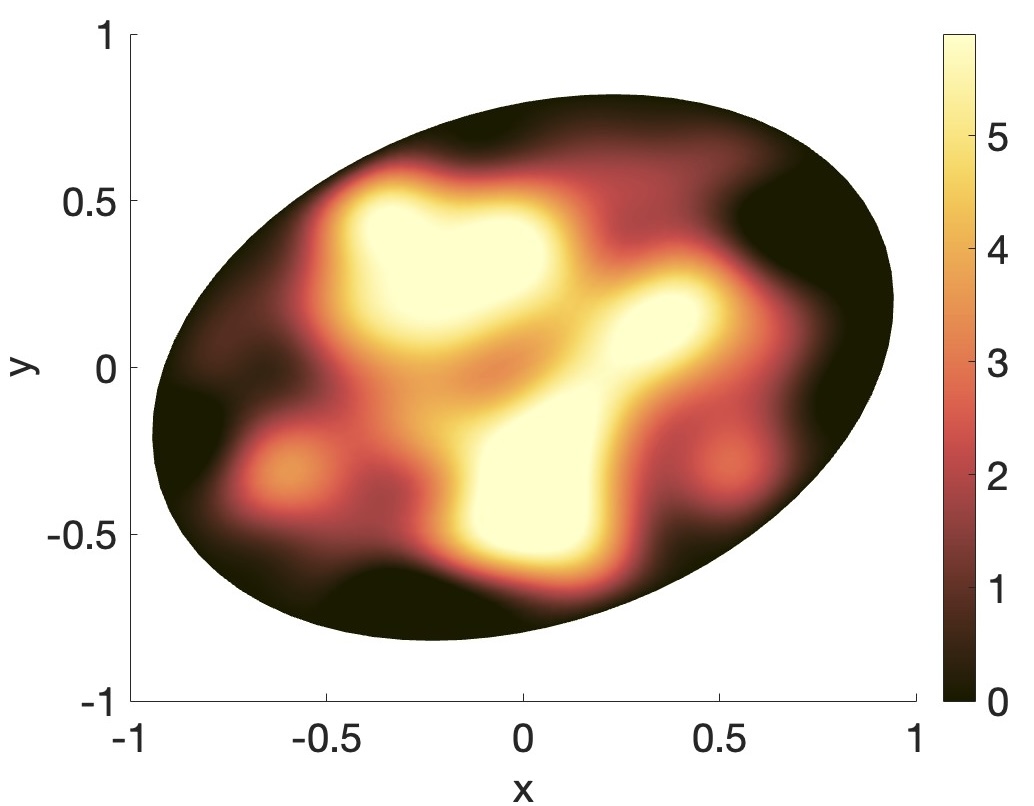}
\caption{Three draws from the Gaussian series prior \eqref{Eq:Prior}, with $\alpha=.5$.}
\label{Fig:Prior}
\end{figure}

	The above prior $\Pi$ can then be be implemented in practice by truncating the series in \eqref{Eq:Prior} at some user-specified level $J\in\N$ (e.g.~at the typical cut-off $J\simeq n^{d/(2\alpha+d)}$ from nonparametric statistics), whereupon the random function $F$ may be identified with the coefficient vector
$$
	\mathbf{F}:= [F_1,\dots,F_J]^T\sim N_J(0,\mathbf{\Lambda}), 
	\qquad \mathbf{\Lambda}:=\text{diag}(\lambda_1^{-\alpha},\dots,\lambda_J^{-\alpha})\in\R^{J,J}.
$$
Three draws from the prior \eqref{Eq:Prior} are depicted in Figure \ref{Fig:Prior}. Similarly discretising the unknown initial condition $f$ by $\mathbf f:=[f_1,\dots,f_J]^T\in\R^J$, with $f_j:=\langle f,e_j\rangle_2$, and recalling the linearity of the forward map $G$, the observation scheme \eqref{Eq:Obs} may be written in matrix notation as
$$
	\mathbf{Y} = \mathbf{G}\mathbf{f} 
	+ \sigma \mathbf W, \qquad \mathbf W:=[W_1,\dots,W_n]^T\sim N_n(0,\mathbf I_n),
$$
where $\mathbf{Y}:=[Y_1,\dots,Y_n]^T$ and $\mathbf{G}:=[Ge_j(x_i), \ j=1,\dots,J,\  i=1,\dots,n]\in \R^{J,n}$. Thus, for given $\mathbf{f}\in\R^J$, $\mathbf{Y}|\mathbf{f}\sim N_n(\mathbf{G}\mathbf{f},\sigma^2\mathbf I_n)$. A standard conjugate computation for multivariate models with Gaussian likelihood and prior then yields
\begin{equation}
\label{Eq:Conjugacy}
	\mathbf f|\mathbf Y\sim N_J(\bar{\mathbf f}_n,\mathbf \Lambda_n),
\end{equation}
with posterior mean and covariance matrix
\begin{equation}
\label{Eq:PostParam}
	\bar{\mathbf f}_n:=\frac{1}{\sigma^2}\mathbf \Lambda_n\mathbf G^T\mathbf Y; 
	\qquad \mathbf \Lambda_n:=\frac{1}{\sigma^2}(\mathbf G^T\mathbf G+\mathbf \Lambda^{-1})^{-1}.
\end{equation}
	
%
%
%

\subsection{Frequentist Asymptotic Properties}\label{Sec:Theory}

In view of the support and regularity properties overviewed above, the prior $\Pi$ in \eqref{Eq:Prior} fits within the theoretical framework of \cite{GN20}, whose result imply the following characterisation of the limiting shape of the posterior distribution $\Pi(\cdot |\mathbf Y)$ as $n\to\infty$, under the assumption that the data $\mathbf Y\sim P_{f_0}^{(n)}$ have been generated according to (an asymptotically equivalent measurement scheme to) model \eqref{Eq:Obs} for some fixed ground truth $f=f_0\in L^2(\Ocal)$. For all sufficiently smooth compactly-supported test functions $\psi\in L^2(\Ocal)$, we have as $n\to\infty$,
\begin{equation}
\label{Eq:BvM}
	\frac{1}{\sqrt n}\left(\langle f,\psi\rangle_2 -  \langle f_0,\psi\rangle_2\right)
	|\mathbf Y \overset{d}{\longrightarrow} N(0,AV(\psi))
\end{equation}
in $P^{(n)}_{f_0}$-probability, where $AV(\psi)$ is the optimal asymptotic variance (in the minimax sense) for estimating the one-dimensional quantity $\langle f,\psi\rangle_2$ from observations $\mathbf Y$ arising as in \eqref{Eq:Obs}; see \cite[Example 3.1]{GK20}.

 	A first important consequence of the semiparametric Berstein-von Mises result \eqref{Eq:BvM} is a central limit for the plug-in posterior mean estimators $\langle \bar f_n,\psi\rangle_2$, where $\bar f_n:=E^\Pi[f|\mathbf Y]$. In particular, it holds that (cf.~\cite[Section 2.2]{GK20})
$$
	\frac{1}{\sqrt n}\left(\langle \bar{f}_n ,\psi\rangle_2 
	- \langle f_0,\psi\rangle_2\right)
	 \overset{d}{\longrightarrow} 
	N(0,AV(\psi))
$$
as $n\to\infty$. In view of the aforementioned minimality of $AV(\psi)$, this implies that $\langle \bar f_n,\psi\rangle_2$ is an asymptotically efficient estimator.

	The second implication concerns the coverage and width of credible intervals built around the efficient estimator $\langle \bar f_n,\psi\rangle_2$,
$$
	C_{n,\gamma}
	:= \{z\in\R:  | z -\langle \bar{f}_n ,\psi\rangle_2|
	\le R_{n,\gamma}\},
	\qquad
	\Pi\left(f:\langle f,\psi\rangle_2\in C_{n,\gamma}|\mathbf Y\right) 
	= 1-\gamma,
$$
where $R_{n,\gamma}>0$ is the appropriate $(1-\gamma/2)\%$-posterior quantile. Under \eqref{Eq:BvM}, it can be shown that $C_{n,\gamma}$ asymptotically achieves the correct frequentist coverage,
$$
	P^{(n)}_{f_0}
	\left( \langle f_0,\psi\rangle_2\in C_{n,\gamma}\right) \to 1-\gamma
$$
as $n\to\infty$, and also that its radius shrinks at optimal rate, $R_{n,\gamma} = O_{P^{(n)}_{f_0}}(n^{-1/2})$. In summary, the (semiparametric) posterior inference resulting from the Gaussian series prior \eqref{Eq:Prior} is valid and optimal from the frequentist point of view. Note that the posterior mean estimator $\bar f_n$ can be readily computed from the explicitly available conjugate formulae \eqref{Eq:PostParam}. Further, although the analytic formulation of the above credible intervals $C_{n,\gamma}$ requires the derivation of the posterior quantiles $R_{n,\gamma}$, these can be efficiently approximated by sampling from the conjugate posterior distribution \eqref{Eq:Conjugacy}.

%
%
%
%
%

\section{Numerical Simulation Study}\label{Sec:Numerics}

\begin{figure}[t]
\centering
\includegraphics[width=3.75cm,height=3.25cm]{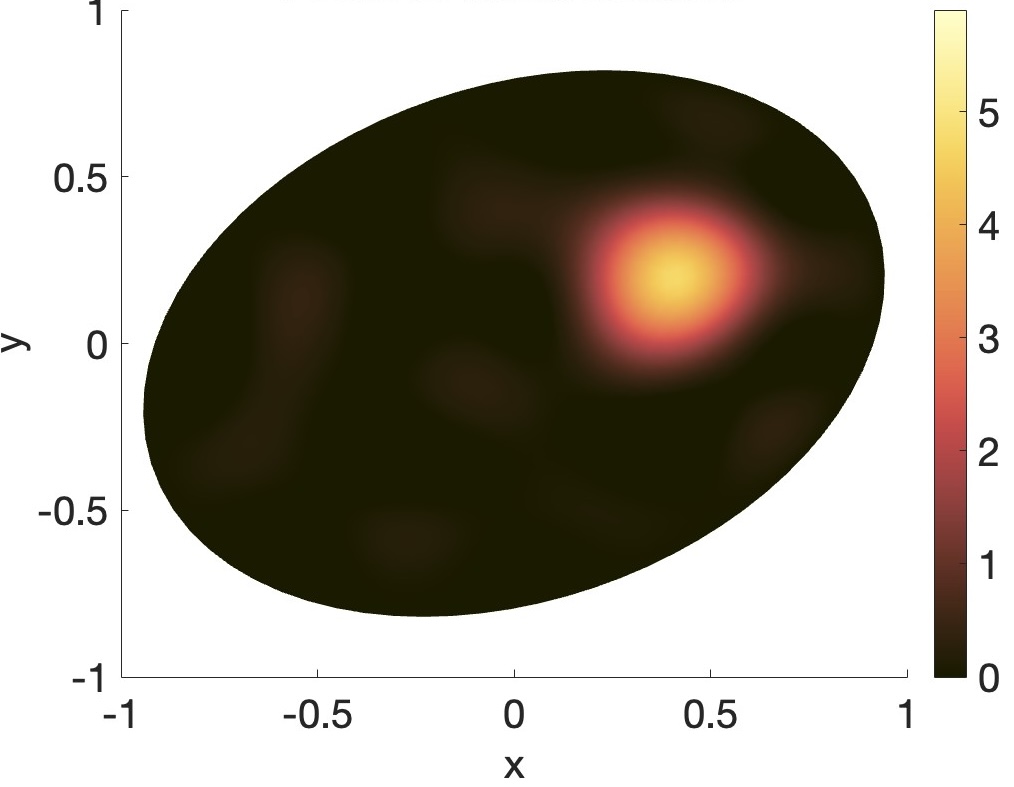}
\includegraphics[width=5.5cm,height=3.25cm]{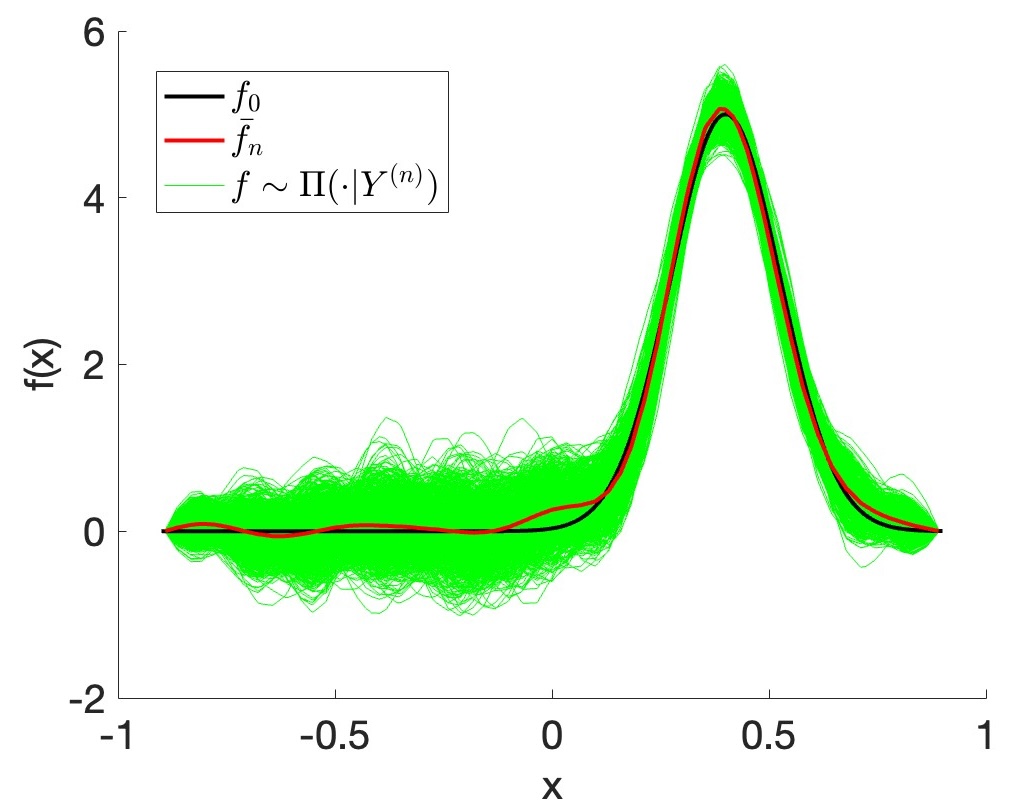}
\caption{Left: the posterior mean estimate $\bar f_n$, computed via the conjugate formulae \eqref{Eq:PostParam}, to be compared with the ground truth $f_0$ depicted in Figure \ref{Fig:StatProb} (left). Right: the cross section of $f_0$ (solid black line), of $\bar f_n$ (red), and of 1000 samples from the posterior distribution (green) along the principal axes of the domain.}\label{fig:HeatEq}
\end{figure}

\begin{table}[t]
\centering
\caption{$L^2$-estimation errors (and relative errors) for the posterior mean estimate $\bar f_n$ of the initial condition, for increasing sample sizes.}
\begin{tabular}{ c|c|c|c|c} 
 	$n$ 			 &   100  & 250   & 500 & 1000\\
 \hline
 $\|\bar f_n - f_0\|_2$ & .3704 (28.8\%) & .3101 (24.8\%) &    .2601 (20.2\%)  &  .2360 (18.3\%)  
\end{tabular}
\label{Tab:Errs}
\end{table}

	We took as the working domain the area $\Ocal$ contained inside a rotated ellipse with horizontal semi-axis of unit length, vertical semi-axis of length $3/4$, and rotation angle $\theta=\pi/6$,
$$
	\{(\cos(t)\cos(\theta) - 3/4\sin(t)\sin(\theta),3/4\sin(t)\cos(\theta)+\cos(t)\sin(\theta)),\
	t\in[0,2\pi)\}.
$$
This choice was made for illustration and other domains could have been considered as well. We fixed the ground truth initial condition $f_0$, cf.~Figure \ref{Fig:StatProb} (left) and eq.~\eqref{Eq:f0}, and generated data $\mathbf Y$ according to \eqref{Eq:Obs} over a deterministic uniform triangular grid $x_1,\dots,x_n\in\Ocal$. The solution $Gf_0$ to the heat equation was numerically computed via finite element methods using MATLAB PDE Toolbox. Given data $\mathbf Y$, we implemented posterior inference with the Gaussian series prior \eqref{Eq:Prior} via the conjugate formulae \eqref{Eq:Conjugacy} and \eqref{Eq:PostParam}. The required discretisation of the forward operator $\mathbf{G}\in\R^{J,n}$ was again computed via MATLAB PDE Toolbox, which was also employed to approximate the Dirichlet-Laplacian eigenpairs (that are not explicitly available for the considered domain $\Ocal$).

Figure \ref{Fig:StatProb} (right) shows $n = 1000$ noisy discrete observations at time $T=.01$ of the solution to the heat equation corresponding to the ground truth
\begin{equation}
\label{Eq:f0}
\begin{split}
	f_0(x,y)
	= 1+e^{-(5x-2)^2-(5y-1)^2},
	\qquad (x,y)\in\Ocal
\end{split}
\end{equation}
(shown above in Figure \ref{Fig:StatProb}, left). The conductivity function was taken to be $s(x,y) = 2.5 - e^{-(5x-2)^2-(2.5y-.5)^2},\ (x,y)\in\Ocal$. The noise standard deviation was set to $\sigma = .05$, yielding a signal-to-noise ratio of $14.43$. The posterior mean estimate $\bar f_n $, computed via \eqref{Eq:PostParam}, is shown in Figure \ref{fig:HeatEq} (left). The $L^2$-estimation error is $\| \bar f_n - f_0\|_2=.2360$. For comparison, $\|f_0\|_2=1.288$, with associated relative error of $18.39\%$. The parameter space was discretised using $J=84$ basis functions. The prior regularity parameter in eq.~\eqref{Eq:Prior} was set to $\alpha=.5$. Table \ref{Tab:Errs} shows the decaying $L^2$-estimation errors for increasing sample sizes $n=100,250,500,1000$, in accordance to the theory overviewed in Section \ref{Sec:Theory}. Figure \ref{fig:HeatEq} (right) shows the cross section of 1000 posterior samples, providing a visualisation of the uncertainty associated to the obtained estimates.


%

\paragraph{Acknowledgement.} This research has been partially supported by MUR, PRIN project 2022CLTYP4.

\bibliographystyle{acm}

\bibliography{SIS25Ref}

\begin{thebibliography}{10}

\bibitem{ASZ14}
{\sc Agapiou, S., Stuart, A.~M., and Zhang, Y.-X.}
\newblock Bayesian posterior contraction rates for linear severely ill-posed
  inverse problems.
\newblock {\em J. Inverse Ill-Posed Probl. 22}, 3 (2014), 297--321.

\bibitem{AMOS19}
{\sc Arridge, S., Maass, P., \"{O}ktem, O., and Sch\"{o}nlieb, C.-B.}
\newblock Solving inverse problems using data-driven models.
\newblock {\em Acta Numer. 28\/} (2019), 1--174.

\bibitem{bissantz2008statistical}
{\sc Bissantz, N., and Holzmann, H.}
\newblock Statistical inference for inverse problems.
\newblock {\em Inverse Probl. 24}, 3 (2008), 034009.

\bibitem{C08}
{\sc Cavalier, L.}
\newblock Nonparametric statistical inverse problems.
\newblock {\em Inverse Probl.}, 24 (2008).

\bibitem{EHN96}
{\sc Engl, H.~W., Hanke, M., and Neubauer, A.}
\newblock {\em Regularization of inverse problems}, vol.~375 of {\em
  Mathematics and its Applications}.
\newblock Kluwer Academic Publishers Group, Dordrecht, 1996.

\bibitem{E10}
{\sc Evans, L.~C.}
\newblock {\em Partial differential equations}, second~ed., vol.~19 of {\em
  Graduate Studies in Mathematics}.
\newblock American Mathematical Society, Providence, RI, 2010.

\bibitem{GvdV17}
{\sc Ghosal, S., and van~der Vaart, A.~W.}
\newblock {\em Fundamentals of Nonparametric Bayesian Inference}.
\newblock Cambridge University Press, New York, 2017.

\bibitem{GK20}
{\sc Giordano, M., and Kekkonen, H.}
\newblock Bernstein--von {M}ises theorems and uncertainty quantification for
  linear inverse problems.
\newblock {\em SIAM/ASA J. Uncertain. Quantif. 8}, 1 (2020), 342--373.

\bibitem{GN20}
{\sc Giordano, M., and Nickl, R.}
\newblock Consistency of {B}ayesian inference with {G}aussian process priors in
  an elliptic inverse problem.
\newblock {\em Inverse Probl. 36}, 8 (2020), 085001--85036.

\bibitem{HT07}
{\sc Haroske, D.~D., and Triebel, H.}
\newblock {\em Distributions, Sobolev Spaces, Elliptic Equations}.
\newblock EMS Press, 2007.

\bibitem{KS04}
{\sc Kaipio, J., and Somersalo, E.}
\newblock {\em Statistical and Computational Inverse Problems}.
\newblock No.~160 in Applied Mathematical Sciences. Springer-Verlag New York,
  2004.

\bibitem{kaltenbacher2005parameter}
{\sc Kaltenbacher, B.}
\newblock Parameter identification in partial differential equations.
\newblock {\em Lecture Notes\/} (2005).

\bibitem{KvdVvZ13}
{\sc Knapik, B.~T., van~der Vaart, A.~W., and van Zanten, J.~H.}
\newblock Bayesian recovery of the initial condition for the heat equation.
\newblock {\em Comm. Statist. Theory Methods 42}, 7 (2013), 1294--1313.

\bibitem{mair1996statistical}
{\sc Mair, B.~A., and Ruymgaart, F.~H.}
\newblock Statistical inverse estimation in hilbert scales.
\newblock {\em SIAM J. Appl. Math. 56}, 5 (1996), 1424--1444.

\bibitem{N23}
{\sc Nickl, R.}
\newblock {\em Bayesian non-linear statistical inverse problems}.
\newblock Zurich Lectures in Advanced Mathematics. EMS Press, Berlin, 2023.

\bibitem{nickl2024consistent}
{\sc Nickl, R.}
\newblock Consistent inference for diffusions from low frequency measurements.
\newblock {\em Ann. Statist. 52}, 2 (2024), 519--549.

\bibitem{S10}
{\sc Stuart, A.~M.}
\newblock Inverse problems: a {B}ayesian perspective.
\newblock {\em Acta Numer. 19\/} (2010), 451--559.

\end{thebibliography}

\end{document}